\documentclass[aps,prd,twocolumn,showpacs,preprintnumbers,superscriptaddress,amsmath,amssymb, longbibliography]{revtex4-1}

\usepackage{bbm}
\usepackage{bm}
\usepackage{color}
\usepackage{dcolumn}
\usepackage{epstopdf}
\usepackage{graphicx}
\usepackage[colorlinks, citecolor=blue]{hyperref}
\usepackage{lipsum}
\usepackage{float}
\usepackage{amsmath}
\usepackage{subfigure}

\newcommand{\fig}[1]{Fig.~\ref{#1}}

\newcommand{\eq}[1]{Eq.~(\ref{#1})}

\newcommand{\beq}{ \begin{equation} }
\newcommand{\eeq}{ \end{equation} }

\newcommand{\e}{\varepsilon}
\newcommand{\s}{\sigma}
\newcommand{\up}{\uparrow}
\newcommand{\down}{\downarrow}

\newcommand{\A}{\mathcal{A}}
\newcommand{\Pol}{\mathcal{P}}

\newcommand{\exch}{\Delta \varepsilon_{\rm exch}}
\newcommand{\exchalpha}{\Delta \varepsilon_{\rm exch}^\alpha}
\newcommand{\exchR}{\Delta \varepsilon_{\rm exch}^R}

\begin{document}

\title{Large voltage-tunable spin valve based on a double quantum dot}

\author{Patrycja Tulewicz}
\email{pattul@st.amu.edu.pl}
\affiliation{Institute of Spintronics and Quantum Information, Faculty of Physics,
	Adam Mickiewicz University, ul. Uniwersytetu Pozna\'nskiego 2, 61-614 Pozna{\'n}, Poland}

\author{Kacper Wrze\'sniewski}
\affiliation{Institute of Spintronics and Quantum Information, Faculty of Physics,
	Adam Mickiewicz University, ul. Uniwersytetu Pozna\'nskiego 2, 61-614 Pozna{\'n}, Poland}

\author{Szabolcs~Csonka}
\affiliation{Department of Physics, Budapest University of Technology and Economics and
	MTA-BME "Momentum" Nanoelectronics Research Group, H-1111 Budapest, Budafoki \'ut 8., Hungary}

\author{Ireneusz Weymann}
\email{weymann@amu.edu.pl}
\affiliation{Institute of Spintronics and Quantum Information, Faculty of Physics,
	Adam Mickiewicz University, ul. Uniwersytetu Pozna\'nskiego 2, 61-614 Pozna{\'n}, Poland}

\date{\today}

\begin{abstract}
We study the spin-dependent transport properties of a spin valve based on a double quantum dot.
Each quantum dot is assumed to be strongly coupled to its own ferromagnetic lead,
while the coupling between the dots is relatively weak.
The current flowing through the system is determined within
the perturbation theory in the hopping between the dots,
whereas the spectrum of a quantum dot-ferromagnetic lead subsystem
is determined by means of the numerical renormalization group method.
The spin-dependent charge fluctuations between ferromagnets and quantum dots
generate an effective exchange field, which splits the double dot levels.
Such field can be controlled, separately for each quantum dot, by the gate voltages
or by changing the magnetic configuration of external leads.
We demonstrate that the considered double quantum dot spin valve setup
exhibits enhanced magnetoresistive properties,
including both normal and inverse tunnel magnetoresistance.
We also show that this system allows for the generation of highly spin-polarized currents,
which can be controlled by purely electrical means. The considered
double quantum dot with ferromagnetic contacts can thus serve
as an efficient voltage-tunable spin valve characterized by high output parameters.
\end{abstract}

\maketitle

\section{Introduction}

Quantum dot spin valves can be regarded as basic building blocks of
quantum spintronics and nanoelectronics \cite{Zutic2004Apr,Awschalom2013Mar}.
Such devices typically consist of a nanoscale object, a quantum dot or a molecule, attached
by tunnel barriers to external ferromagnetic contacts \cite{Barnas1998Feb,Rudzinski2001Aug,
	Pasupathy2004Oct,Sahoo2005Nov,Samm2014May}.
The current flowing through these systems
strongly depends on the mutual orientation of the magnetic moments
of ferromagnetic leads and can be additionally controlled by a gate
voltage. In conventional spin valves, the tunneling current
is larger when the configuration of the leads' magnetizations
is parallel as compared to the case when the magnetic moments
of the leads form an antiparallel configuration \cite{Barnas2008Sep}. This is associated with
the asymmetry between the spin-dependent tunneling  processes
in these two configurations \cite{Julliere1975Sep,Maekawa2006Mar}. The situation becomes more interesting
when strong electron correlations are present in the system,
such as the ones driving the Kondo effect \cite{Kondo1964,Hewson1997,Goldhaber1998}.
Then, an inverse tunnel magnetoresistance effect may develop in the system \cite{Pasupathy2004Oct,Hamaya2008Feb,Weymann2011Mar}.
Moreover, the current flowing through the device can become highly spin-polarized \cite{Hamaya2007Jan,Merchant2008Apr}.
In fact, a single quantum dot based spin polarizer,
with nearly perfect electrically-controlled spin polarization of the tunneling current,
has been recently proposed \cite{Csonka2012May}. The operation of such device is based
on the exchange field that is induced when quantum dot or molecule
becomes strongly coupled to ferromagnetic reservoirs \cite{Martinek2003,Lopez2003Mar,Martinek2005}.
By spin-dependent charge fluctuations, the orbital level of the dot
becomes spin-split and consequently mainly the electrons with favorable spin orientation
are transferred through the system. Moreover, it turns out
that the exchange field results in high spin polarization
of the local density of states, which is responsible for
controllable spin rectification of the current \cite{Csonka2012May}

From a theoretical point of view, the transport properties of quantum dot spin valves,
consisting of single dots coupled to ferromagnetic leads,
have already been a subject of extensive studies, both in the weak
\cite{Barnas1998Feb,Rudzinski2001Aug,Konig2003Apr,Braun2004Nov,Weymann2005Sep,Barnas2008Sep}
and in the strong coupling regime \cite{Lopez2003Mar,Martinek2003,Martinek2005,Weymann2011Mar,Gergs2018Jan}. Moreover,
more complex structures involving double quantum dots
with ferromagnetic leads have also been explored \cite{Weymann2008Jul,Zitko2012Apr,Wojcik2015Apr,Weymann2018Feb}.
As far as the experimental progress is concerned,
most experiments on nanoscopic spin valves
have been devoted to transport properties of single dot or molecular
structures \cite{Pasupathy2004Oct,Sahoo2005Nov,Hamaya2007Jul,Hamaya2007Dec,Hamaya2008Feb,
	Hauptmann2008Mar,Merchant2008Apr,Gaass2011Oct,Samm2014May,Dirnaichner2015May},
whereas double quantum dot spin valves
have been implemented only very recently \cite{Bordoloi2020Aug}.

\begin{figure}[t]
	\includegraphics[width=1\columnwidth]{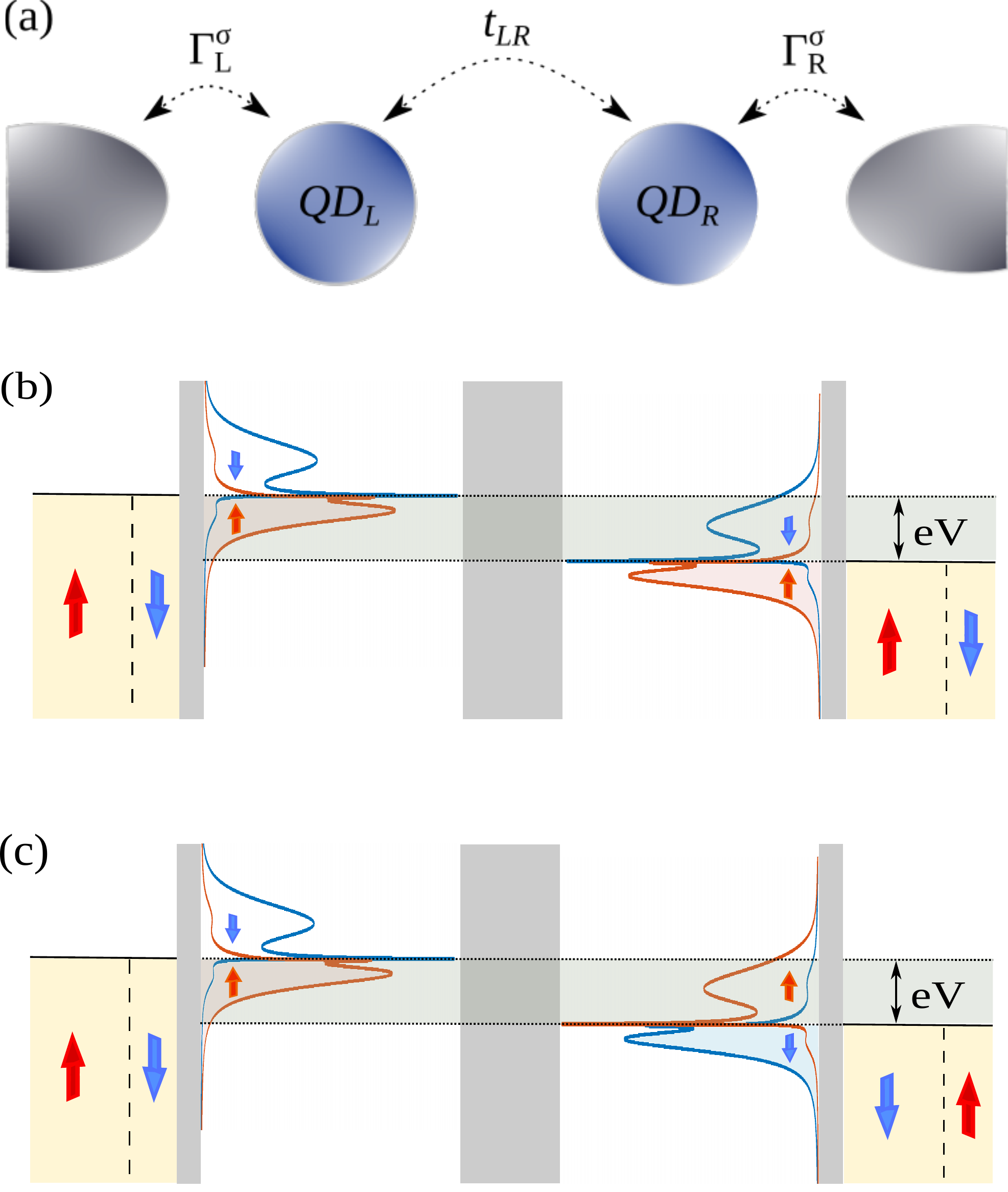}
	\caption{\label{Fig:1}
		(a) Schematic of the considered system.
		It consists of two quantum dots: the left (\(\alpha=L\)) and right (\(\alpha=R\)) one,
		attached to the corresponding metallic ferromagnetic leads
		with the spin-dependent coupling  strengths $\Gamma_{\alpha}^\sigma$.
		The coupling between the dots is represented by the hopping matrix element $t_{LR}$.
		The diagrams for the parallel and antiparallel configuration of leads'
		magnetic moments, demonstrating the spin valve behavior,
		are presented in (b) and (c), respectively.
		The quantum dots are assumed to be occupied by a single electron,
		such that the density of states is characterized by large spin asymmetry due to the exchange field.
		The magnitude and spin polarization of the current flowing through the system
		is determined by the polarization of the local density of states in the bias window $eV$.
		Due to the mismatch of densities of states, the current in the parallel
		configuration is much smaller compared to the current flowing in
		the antiparallel configuration, which is additionally highly spin polarized.
	}
\end{figure}

In this paper we develop the theory of transport through a double quantum dot
strongly coupled to external ferromagnetic leads, while the coupling
between the two dots is assumed to be relatively weak, see \fig{Fig:1}(a).
In such geometry, the spin-resolved transport
strongly depends on the magnitude and sign of
the exchange field present on each quantum dot.
In fact, such field plays a role of a local magnetic field
that can be controlled by the gate voltage applied to the dot.
The possibility of splitting the level of a given dot at will provides
the opportunity for implementing a spin valve with
magnetoresistive response much exceeding that
obtained for external-magnetic-field controlled behavior \cite{Bordoloi2020Aug}.
We show that the particular geometry of the system
considered in this work allows us for obtaining enhanced spin valve
behavior, ensuring full control of spin-dependent transport by the bias and gate voltages.
In particular, in the linear response regime, we demonstrate
that one can obtain a considerable, both positive and inverse, tunnel magnetoresistance.
On the other hand, in the nonlinear response regime,
we predict a perfect spin polarization of the flowing current
and a greatly enhanced magnetoresistance when
the sign of the exchange field is different in each dot,
which can be obtained by appropriate tuning of gate voltages.
Our work reveals thus that double quantum dots strongly attached to
external ferromagnetic leads can be regarded as fully voltage-controllable
spin valves with very prospective spin-resolved properties.

This paper is structured as follows. Section II is devoted
to theoretical description, where the model, method and formulas
for transport quantities are presented. In Sec. III we discuss the
possibility of gate-voltage control of double-dot
spin valve in the linear response regime,
whereas Sec. IV is devoted to the spin-resolved
transport properties when a finite bias voltage
is applied to the system. Finally, the paper is concluded in Sec. V.


\section{Theoretical description} \label{theor_disc}


The schematic of the considered double quantum dot spin valve is presented in \fig{Fig:1}(a).
The system consists of two single-level quantum dots
strongly attached to external ferromagnetic leads with coupling strengths
$\Gamma_{\alpha}^{\sigma}$, where $\alpha = L$ ($\alpha = R$)
for the left (right) lead, and weakly connected to each other
through the hopping matrix elements $t_{LR}$.
The ferromagnetic leads are assumed to form either parallel
or antiparallel magnetic configuration.
The total Hamiltonian of the system can be written as
\begin{equation}
	H = H_{0L} + H_{0R} + H_{TL} + H_{TR} + H_{DQL}  + H_{DQR}+ H_{T},
	\label{Eq:H}
\end{equation}
where $H_{0\alpha}$ models the noninteracting electrons
in the left ($\alpha=L$) and right ($\alpha=R$) ferromagnetic electrode,
\begin{equation}
  H_{0\alpha} = \sum_{k \sigma}\varepsilon_{\alpha k\sigma}{c_{\alpha k\sigma}^\dagger}c_{\alpha k\sigma},
  \label{eq:4}
\end{equation}
where the operator $c_{\alpha k\sigma}^\dagger$ creates an electron with momentum $k$
spin $\sigma$ of energy $\varepsilon_{\alpha k\sigma}$ in the lead $\alpha$.
The tunneling between the corresponding quantum dot and electrode $\alpha$ is modeled by
\begin{equation}
	H_{T\alpha}=\sum_{k \sigma} t_{\alpha}^\sigma (c_{\alpha k\sigma}^\dagger d_{\alpha\sigma} + d_{\alpha\sigma}^\dagger c_{\alpha k\sigma}),
	\label{eq:5}
\end{equation}
with $d_{\alpha\sigma}^\dag$ ($d_{\alpha\sigma}$) being the creation (annihilation)
operator of a spin-$\sigma$ electron in the quantum dot $\alpha$
and $t_{\alpha}^\sigma$ denoting the tunnel matrix elements, assumed to be momentum independent.
The strength of the tunnel coupling between the corresponding quantum dot and electrode
can be written as $\Gamma_\alpha^\sigma = \pi\rho_\alpha^\sigma |t_\alpha^\sigma|^2$,
where $\rho_\alpha^\sigma$ is the spin-dependent density of states
of lead $\alpha$. The coupling constants can be more conveniently written
in terms of spin polarization of lead $\alpha$,
$p_\alpha = (\Gamma_\alpha^\uparrow - \Gamma_\alpha^\downarrow ) /
(\Gamma_\alpha^\uparrow + \Gamma_\alpha^\downarrow )$, as
$\Gamma_\alpha^\sigma = (1+\hat{\sigma}p_\alpha)\Gamma_\alpha$,
where $\hat{\sigma} = +(-)$  for majority (minority) spin band
and $\Gamma_\alpha=(\Gamma_\alpha^\uparrow + \Gamma_\alpha^\downarrow )/2$.
The hopping between the two quantum dots is described by
\begin{equation}
	H_{T}=\sum_{\sigma} t_{LR} (d_{L \sigma}^\dagger d_{R\sigma} + d_{R\sigma}^\dagger d_{L\sigma}).
	\label{eq:6}
\end{equation}
Finally, the Hamiltonian for quantum dot $\alpha$ is given by
\begin{equation}
		H_{DQ\alpha} = \sum_{\s} \e_\alpha n_{\alpha\s} +  U_\alpha n_{\alpha\uparrow } n_{\alpha\downarrow } ,
		\label{Eq:HDQD}
\end{equation}
where $n_\alpha = n_{\alpha\uparrow} + n_{\alpha\downarrow}$
with $n_{\alpha\sigma} = d_{\alpha\sigma}^\dag d_{\alpha\sigma}$.
The Coulomb correlations on each quantum dot are denoted by $U_\alpha$.
In our considerations, it is assumed that the coupling between the two dots
is much weaker than the coupling of each dot to its contact.
Moreover, we also assume that the on-site Coulomb correlations
are much larger than the correlations between the two dots
and therefore the latter ones can be neglected.

To determine the current flowing through the system,
we perform a perturbative expansion in $H_{T}$.
Consequently, we rewrite the Hamiltonian (\ref{Eq:H}) as,
$H=H_L + H_R + H_{T}$, where $H_\alpha $ models
the dot $\alpha$ coupled to the corresponding lead.
It is given by the Anderson Hamiltonian \cite{Anderson1961r},
$H_\alpha = H_{0\alpha } + H_{T \alpha } + H_{QD\alpha}$.
Then, in the lowest-order with respect to hopping matrix elements $t_{LR}$,
we can express the current flowing in the spin channel $\sigma$ as \cite{Nazarov2009May,Csonka2012May}
\begin{eqnarray}
	I_\sigma(V)&=&\frac{1}{eR_\sigma}
	\int_{-\infty}^{\infty} \mathrm{d}\omega \A_{L}^{\sigma}(\omega) \A_{R}^{\sigma}(\omega-eV)\nonumber\\
	&&\times [f(\omega)-f(\omega-eV)],
	\label{Eq:I}
\end{eqnarray}
where $f(\omega)$ is the Fermi-Dirac distribution function and
the resistance of the junction between the two quantum dots is given by,
$R_\sigma = \hbar / (2\pi e^2 A_{L0}^\sigma A_{R0}^\sigma t_{LR}^2)$,
with $A_{\alpha0}^\sigma = 1/\pi\Gamma_\alpha^\sigma$.
$\A_{\alpha}^\sigma(\omega)$ is the normalized spin-dependent spectral function
(local density of states) of the quantum dot $\alpha$ coupled to the corresponding lead, as described by $H_\alpha$.
Note that we assume the left lead to be grounded, while
the voltage is applied to the right lead, see \fig{Fig:1}.
In the following, we assume that the quantum dots have comparable charging
energies and are coupled with the same strength to the leads,
i.e. $U_L = U_R \equiv U$ and $\Gamma_L = \Gamma_R \equiv \Gamma$.
We also set $t_{LR} = \Gamma/10$.
Moreover, we assume that the electrodes are made of the same material, such that $p_L=p_R=p$.
Taking into account realistic quantum dot parameters,
$\Gamma \approx 1$meV and $t_{LR}\approx 0.1$meV,
one finds $R_\s \sim {\rm M}\Omega$, which yields the maximum current on the order of tens of nA.

In this work we are interested in the spin-dependent
transport properties of the system at low temperatures
both in the linear and nonlinear response regimes.
The main computational task is to determine the spin-resolved spectral function
$A_{\alpha}^\sigma(\omega)$ of the quantum dot-ferromagnetic lead subsystem modeled by $H_\alpha$.
The spectral function is given by, $A_{\alpha}^\sigma(\omega) = -{\rm Im } [G_{\alpha}^\sigma(\omega)]/\pi $,
where $G_{\alpha}^\sigma(\omega)$ is the Fourier transform
of the retarded Green's function of the corresponding quantum dot level,
$G_{\alpha}^\sigma(t) = -i\Theta(t) \langle  \{ d_{\alpha\s}^\dag (0), d_{\alpha \s}(t) \}\rangle$.
Because each dot is strongly coupled to its own lead, while
the hopping between the two dots is assumed to be the smallest energy scale,
we can use the numerical renormalization group (NRG) method \cite{Wilson1975,NRG_code,Bulla2008}
to find $A_{\alpha}^\sigma(\omega)$ separately for each quantum dot.
This method allows us to accurately account for all the correlation
effects between the quantum dot and ferromagnet.
In calculations, we use the discretization parameter $\Lambda=2$
and keep at least $N_K=1024$ low-energy states in the numerical procedure.
Moreover, to increase the accuracy of spectral functions,
we employ the $z$-averaging trick \cite{Oliveira_PhysRevB.41.9403} and make use of
the optimal broadening method \cite{Florens_PhysRevB.79.121102}.

For nonmagnetic leads, at sufficiently low temperatures,
the quantum dot-electrode subsystem exhibits the Kondo effect \cite{Kondo1964,Goldhaber1998}.
On the other hand, for ferromagnetic leads, the spin-resolved charge
fluctuations between the dot and ferromagnet give rise to an effective exchange field $\exch$ \cite{Martinek2003}.
Such field can spin-split the quantum dot level and, consequently,
suppress the Kondo effect if the splitting becomes larger than the corresponding
Kondo temperature \cite{Hamaya2007Dec, Hauptmann2008Mar, Gaass2011Oct}.
Consequently, the exchange field plays a role of a highly-tunable local magnetic field acting on a given dot.
Analytically, this field can be estimated within the second-order perturbation theory,
and at zero temperature, for quantum dot $\alpha$,
it is given by \cite{Martinek2003,Martinek2005}
\begin{equation}
	\exchalpha = \frac{2\Gamma  p} { \pi} {\rm ln}\Big| \frac{\e_\alpha} {\e_\alpha+U} \Big|.
\end{equation}
It is evident that the field changes sign when the dot level
crosses the particle-hole symmetry point, $\e_\alpha = -U/2$.
Moreover, the exchange field is responsible for high spin polarization
of the spectral functions \cite{Csonka2012May}. This can be seen in Figs. \ref{Fig:1}(b) and (c),
which demonstrate the principle of operation of the
considered double quantum dot spin valve in the nonequilibrium regime.
The presented spectral functions were calculated by NRG for $\e_L = \e_R = -0.3 U$,
i.e. when $\exchalpha$ is finite for each dot,
and clearly reveal a strong spin asymmetry.
Note also the presence of a small resonance due to the Kondo effect,
which is pinned to the lead's chemical potential.
The large spin valve behavior can be obtained either by
tuning the positions of the quantum dot levels, which
can be done with appropriate gate voltages
or by changing the magnetic configuration of the device.
The latter mechanism is sketched in Figs. \ref{Fig:1}(b) and (c).


\section{Linear response regime} \label{lin}

In this section we discuss the linear-response behavior of the system.
The linear conductance in the spin channel $\sigma$ can be found from
\begin{equation}
	G_\sigma=\frac{1}{R_\sigma}
	\int_{-\infty}^{\infty} \mathrm{d}\omega \A_{L}^{\sigma}(\omega) \A_{R}^{\sigma}(\omega)
	\left(\! -\frac{\partial f(\omega)}{\partial\omega} \right).
	\label{Eq:Glin}
\end{equation}
As results from the above formula, for low temperatures
the linear conductance is proportional to the product
of the two spectral functions at the Fermi level.
Thus, appropriate tuning of the quantum dot levels,
which can be experimentally achieved by gate voltages, can greatly
affect the conductance of the system.
Moreover, it will determine the spin-resolved properties
of the considered spin valve, allowing for an accurate
tuning of both the tunnel magnetoresistance and the spin polarization
of the current in a fully electrical manner, without the need
for applying external magnetic field. Nevertheless,
besides such electric control, changing the magnetic configuration
of the device will be another way to tune the spin-dependent behavior.

\begin{figure}[t]
	\includegraphics[width=0.8\columnwidth]{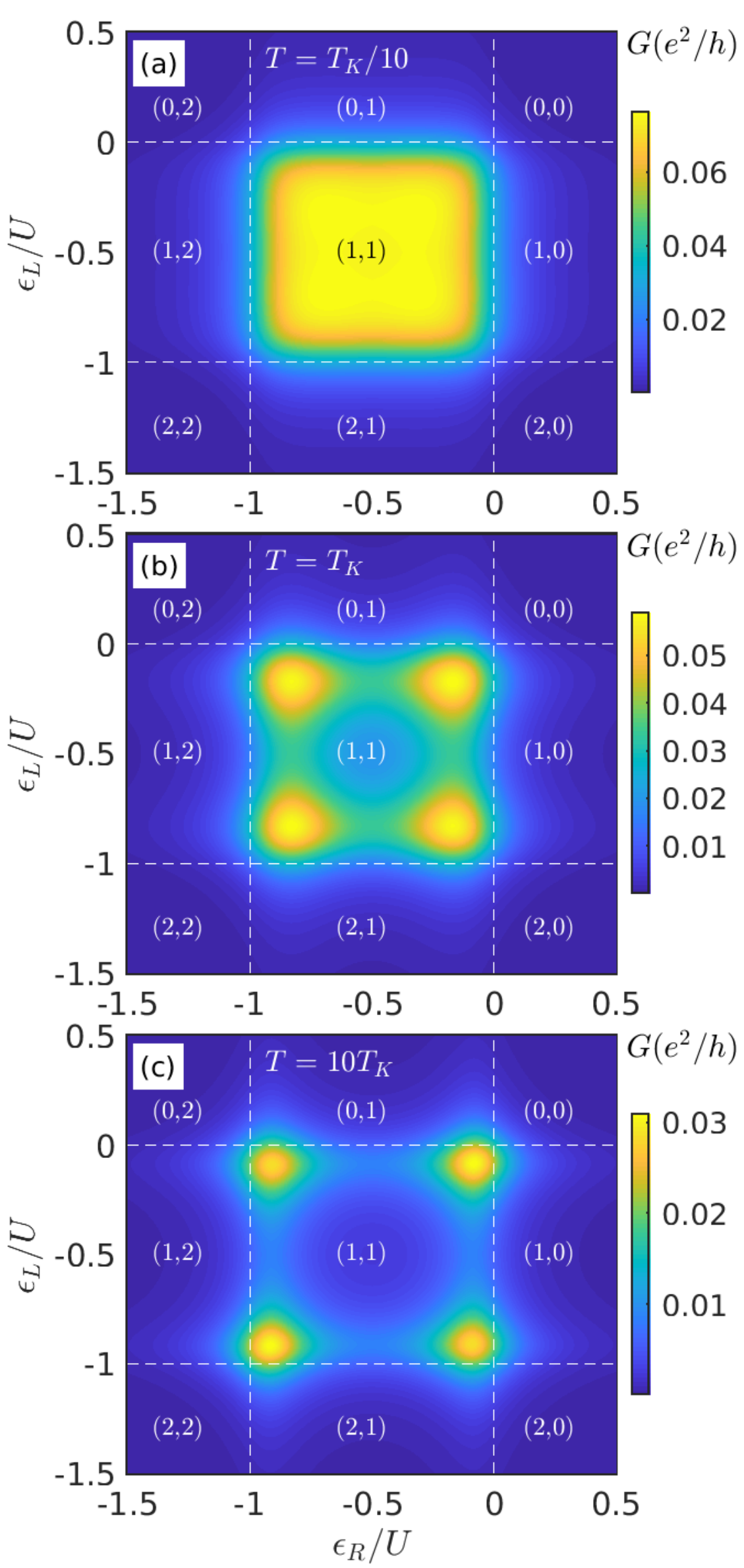}
	\caption{\label{Fig:NM2D}
		The linear conductance $G$ for nonmagnetic system calculated as a function of
		the position of each quantum dot level $\varepsilon_L$ and $\varepsilon_R$
		for different temperatures: (a) $T=T_K/10$, (b)  $T=T_K$, (c)  $T=10T_K$.
		The numbers in brackets indicate
		the occupation of each quantum dot, \(( \langle n_L \rangle,\langle n_R \rangle)\),
		while the dashed lines separate the regions with different occupations.
		The parameters are: $U = 0.2$, $\Gamma=0.015$ and $t_{LR}=\Gamma/10$
		in units of band halfwidth, and $p=0$. The Kondo temperature
		is $T_K\approx 0.0002$.
	}
\end{figure}

To begin with, in \fig{Fig:NM2D} we present the linear conductance
$G = \sum_\s G_\sigma$ as a function of quantum dot energy levels, $\e_L$ and $\e_R$,
in the case of nonmagnetic leads. We note that since the
level positions can be changed by gate voltages, this figure
effectively presents the gate voltage dependence of the conductance.
The dashed lines in the figure mark the regions
with different charges in the double dot, as indicated.
This figure was determined for three different temperatures,
$T=T_K/10$, $T=T_K$ and $T=10T_K$,
where $T_K$ is the Kondo temperature of a single dot in the case of
$\e_\alpha  =-U/2$ and $p=0$, which is approximately equal to
\cite{Haldane_PhysRevLett.40.416} $T_K\approx 0.0002$
in units of band halfwidth $D$, henceforth used as the energy unit $D\equiv 1$.

It is easy to understand the low values of conductance in the transport regions
with even occupation on each quantum dot. Then, the Kondo effect between
the dot and the corresponding lead does not develop and the spectral functions
have relatively low weights, which effectively results in a suppressed conductance.
When one of the dots is oddly occupied, it hosts the Kondo effect
characterized by a pronounced resonance in the spectral function at the Fermi level.
For single quantum dots coupled to the leads this would give rise to the unitary conductance through the system \cite{Hewson1997}.
However, in the considered setup this is clearly not the case,
as the Kondo effect on one of the dots does not result
in large conductance through the whole system
due to much lower spectral weight in the other dot.
Only when the two quantum dots exhibit the Kondo effect at the same time,
i.e. in the charge sector $(1,1)$, the conductance through
the system may become considerable.
This is clearly visible
in \fig{Fig:NM2D}(a), which was calculated for $T<T_K$.
For temperatures much lower than $T_K$, there is a plateau of relatively large $G$ for
$-U\lesssim \e_L \lesssim 0$ and $-U\lesssim \e_R \lesssim 0$.
This plateau is associated with the fact that both dots exhibit
the Kondo effect in the single occupied regime.
When the temperature is increased such that
$T = T_K$ (note that $T_K$ is estimated for $\e=-U/2$),
one observes a gradual decrease of $G$,
especially in the middle of the $(1,1)$ Coulomb blockade region, see \fig{Fig:NM2D}(b).
On the other hand, when $T>T_K$, the suppression becomes larger, such that only
local maxima at the corners of the $(1,1)$ occupancy region
are present, see \fig{Fig:NM2D}(c).
Assuming vanishing temperature and that each quantum dot
exhibits the Kondo effect, the linear conductance is given by,
\begin{equation} \label{Eq:GNM}
	G \approx \frac{e^2}{h} \frac{8t_{LR}^2}{\Gamma^2}.
\end{equation}
Note that the maximum value of the conductance is conditioned here by the
value of the hopping between the dots, which is assumed to be much smaller than the coupling to
external leads. This is why, although the conductance exhibits a
plateau for $T\ll T_K$, its maximum value
is still considerably lower than the unitary conductance.

\begin{figure}[t]
	\includegraphics[width=1\columnwidth]{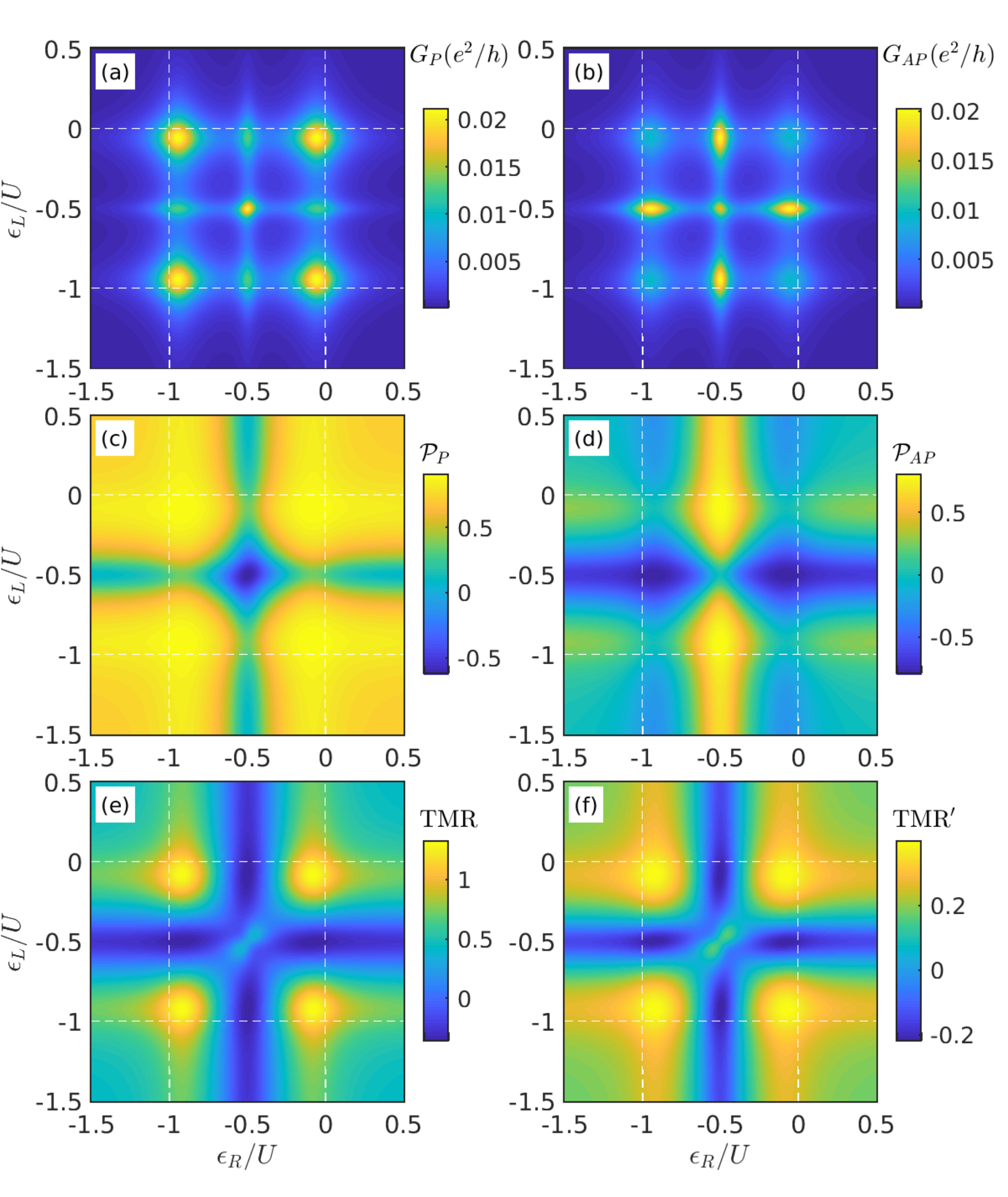}
	\caption{\label{Fig:FM2DTK}
		(a,b) The linear conductance $G_X$ and (c,d) current spin polarization $\Pol_X$
		in the case of (a,c) parallel ($X=P$) and (b,d) antiparallel ($X=AP$)
		magnetic configuration as well as the resulting tunnel magnetoresistance
		(e) TMR and (f) ${\rm TMR}^\prime$ as a function of $\varepsilon_L$ and $\varepsilon_R$
		for $T=T_K$.
		The dashed lines separate the regions with different occupations, as marked in \fig{Fig:NM2D}.
		The parameters are the same as in \fig{Fig:NM2D} with $p=0.4$.
	}
\end{figure}

Having presented the main features of the conductance in the nonmagnetic case, we
are now ready to study the situation when the contacts are magnetic.
We assume that the magnetizations of the leads can form
two magnetic configurations, the parallel ($P$) and antiparallel ($AP$) one \cite{Pasupathy2004Oct,Sahoo2005Nov}.
Having determined the linear conductance in
these two magnetic configurations, $G_P$ and $G_{AP}$, we can calculate
the tunneling magnetoresistance (TMR) of the device, which is defined as \cite{Julliere1975Sep}
\begin{equation}
	{\rm TMR} = \frac{G_P - G_{AP}}{G_{AP}}.
	\label{eq:TMR}
\end{equation}
We note that the TMR can be also defined in a different manner \cite{Bordoloi2020Aug}
\begin{equation}
    {\rm TMR}^\prime = \frac{G_P - G_{AP}}{G_{P} + G_{AP}}.
	\label{eq:TMR2}
\end{equation}
Because the two definitions result in different values of tunnel magnetoresistance
and it is not easy to infer the value of e.g. ${\rm TMR}^\prime$
from the knowledge of TMR, in the following, to make the picture complete,
we present the both cases. Another important quantity characterizing the spin-resolved
transport properties of the device is the spin polarization of the current.
In the linear response regime it can be defined as
\begin{equation}
	\Pol_{X} = \frac{G_\uparrow^X - G_\downarrow^X}{G_\uparrow^X + G_\downarrow^X},
	\label{eq:Pol}
\end{equation}
where $G_\s^X$ is the linear conductance in the spin channel $\sigma$
in the case of parallel $(X=P)$ or antiparallel $(X=AP)$ magnetic configuration.

The linear response conductance and spin polarization of the current in both magnetic configurations,
as well as the tunnel magnetoresistance calculated
as a function of $\e_L$ and $\e_R$ at temperature $T=T_K$ are presented in \fig{Fig:FM2DTK}.
We note that in the case of magnetic contacts
the spin-resolved transport properties only weakly change for $T\lesssim T_K$,
therefore we do not show the data for lower temperatures here.
In fact, when $T \ll T_K$, it is only the conductance exactly for $\e_L=\e_R=-U/2$
that becomes enhanced, while the other properties are hardly affected.
This is associated with the fact that for $\e_L=\e_R=-U/2$
the exchange fields cancel and both quantum dot local densities of states exhibit the Kondo resonance,
resulting in an enhancement of linear conductance with lowering the temperature.
At zero temperature, the conductance in the parallel and antiparallel magnetic
configuration for $\e_L=\e_R=-U/2$ can be approximated by
\begin{eqnarray}
	G_{P} &\approx& \frac{e^2}{h}\frac{8t_{LR}^2}{\Gamma^2}\frac{1+p^2}{(1-p^2)^2},
		\label{Eq:GP}\\
	G_{AP} &\approx& \frac{e^2}{h}\frac{8t_{LR}^2}{\Gamma^2}\frac{1}{1-p^2}	.
	\label{Eq:GAP}
\end{eqnarray}
It is interesting to note that for $T\ll T_K$ both conductances are larger by a spin-polarization-dependent
factor, as compared to the case of nonmagnetic leads, cf. \eq{Eq:GNM}.
This is however only the case exactly in the middle of the $(1,1)$ occupancy regime,
since detuning from this point results in an immediate and large
suppression of $G$ due to the spin splitting of the orbital levels
by the exchange field, see Figs.~\ref{Fig:FM2DTK}(a) and (b).
In fact, the exchange field in each quantum dot,
$\exchalpha$, plays a role of a local magnetic field that acts only on
appropriate dot. Moreover, the magnitude and sign of this field
can be controlled by purely electrical means, i.e. by tuning the gate voltages.
This is clearly an asset of the setup discussed here
over the magnetic field controlled spin valve
demonstrated in Ref.~\cite{Bordoloi2020Aug}.

Using the above formulas for the conductance in the case of $\e_L=\e_R=-U/2$,
one can find the tunnel magnetoresistance,
${\rm TMR} \approx 2 p^2 / (1-p^2)$ and ${\rm TMR}^\prime \approx p^2$.
On the other hand, the spin polarization in both
magnetic configurations is $\Pol_P \approx -2p/(1+p^2)$ and $\Pol_{AP} \approx 0$.
It is interesting to note that the value of ${\rm TMR}$ for $\e_L=\e_R=-U/2$
is the same as that predicted by the Julliere model
for a single magnetic tunnel junction \cite{Julliere1975Sep}.
However, when the system is detuned from this point,
the spin-resolved behavior becomes greatly changed.
One can then observe both enhanced and inverse TMR effect
as well as full current spin-polarization,
which demonstrates a clear advantage of the considered DQD setup
over a simple tunnel junction, as we show in the sequel.

When the system is detuned from the particle-hole symmetry point of each dot $\e_L=\e_R=-U/2$,
the exchange field starts playing an important role
and determines the system's transport properties.
First of all, one can see that the conductance becomes then
generally suppressed and it only shows lines of enhanced
conductance in the $(1,1)$ region for either
$\e_L=-U/2$ or $\e_R=-U/2$, see Figs.~\ref{Fig:FM2DTK}(a) and (b). For these values of the level position,
there is a Kondo resonance in the local density of states of one of the dots.
Consequently, the cross-like feature visible in the middle
Coulomb blockade region has the Kondo origin.
In addition, there are also local maxima at the corners
of the $(1,1)$ occupancy region, similarly as in the nonmagnetic case,
cf. \fig{Fig:NM2D}(b). These are related with enhanced tunneling processes
when the dots are on resonance either for $\e_\alpha\approx 0$
or $\e_\alpha\approx -U$. The above described behavior
is present in both magnetic configurations. However,
importantly, the magnitude of those effects is completely different,
resulting  in either positive or negative tunnel magnetoresistance,
see Figs.~\ref{Fig:FM2DTK}(e) and (f). As far as the
Kondo cross-like pattern is concerned, one can see that
the  TMR is negative in this parameter space, signaling that $G_{AP} > G_{P}$.
This can be understood by realizing that when one of the dots is at
half-filling its local density of states at low temperatures
is given by $A_{0\alpha }^\sigma = 1/\pi\Gamma_{\alpha}^\sigma$.
Assume that $\e_L=-U/2$, then for the linear conductance at $T=0$ one gets
\begin{eqnarray}
G_{P/AP} &\approx& \frac{e^2}{h}\frac{4\pi t_{LR}^2}{(1-p^2)\Gamma}\left[A_{R}^{\uparrow}(0)+A_{R}^{\downarrow}(0)\right.\nonumber\\
&& \left.\mp p(A_{R}^{\uparrow}(0)-A_{R}^{\downarrow}(0))  \right].
\end{eqnarray}
Note that $A_{\alpha}^{\sigma}(0)$ is the spectral function of dot $\alpha$
for spin $\sigma$ taken at the Fermi energy, which depends on the position
of the dot level.
Since, generally, $A_{R}^{\uparrow}(0) >  A_{R}^{\downarrow}(0)$
due to the strong coupling to the spin-up channel (except for $\e_L=\e_R=-U/2$),
one indeed finds that $G_{AP}>G_P$, resulting in inverse TMR effect.
On the other hand, large values of conductance at the corners
of the  $(1,1)$ blockade region in the parallel magnetic configuration
as compared to the antiparallel one are associated with enhanced
transport properties in the majority spin channel.
More specifically, in the parallel alignment the
conductance is determined by a product of the local densities of states
for the spin-up component which give the dominant contribution.
In the antiparallel alignment, in turn, $G_{AP}$ is always
proportional to a product of the spin majority and spin minority
densities of states. As a consequence, transport is more effective
in the parallel configuration, $G_P>G_{AP}$,
giving rise to positive TMR, see Figs.~\ref{Fig:FM2DTK}(e) and (f).

\begin{figure}[t]
	\includegraphics[width=1\columnwidth]{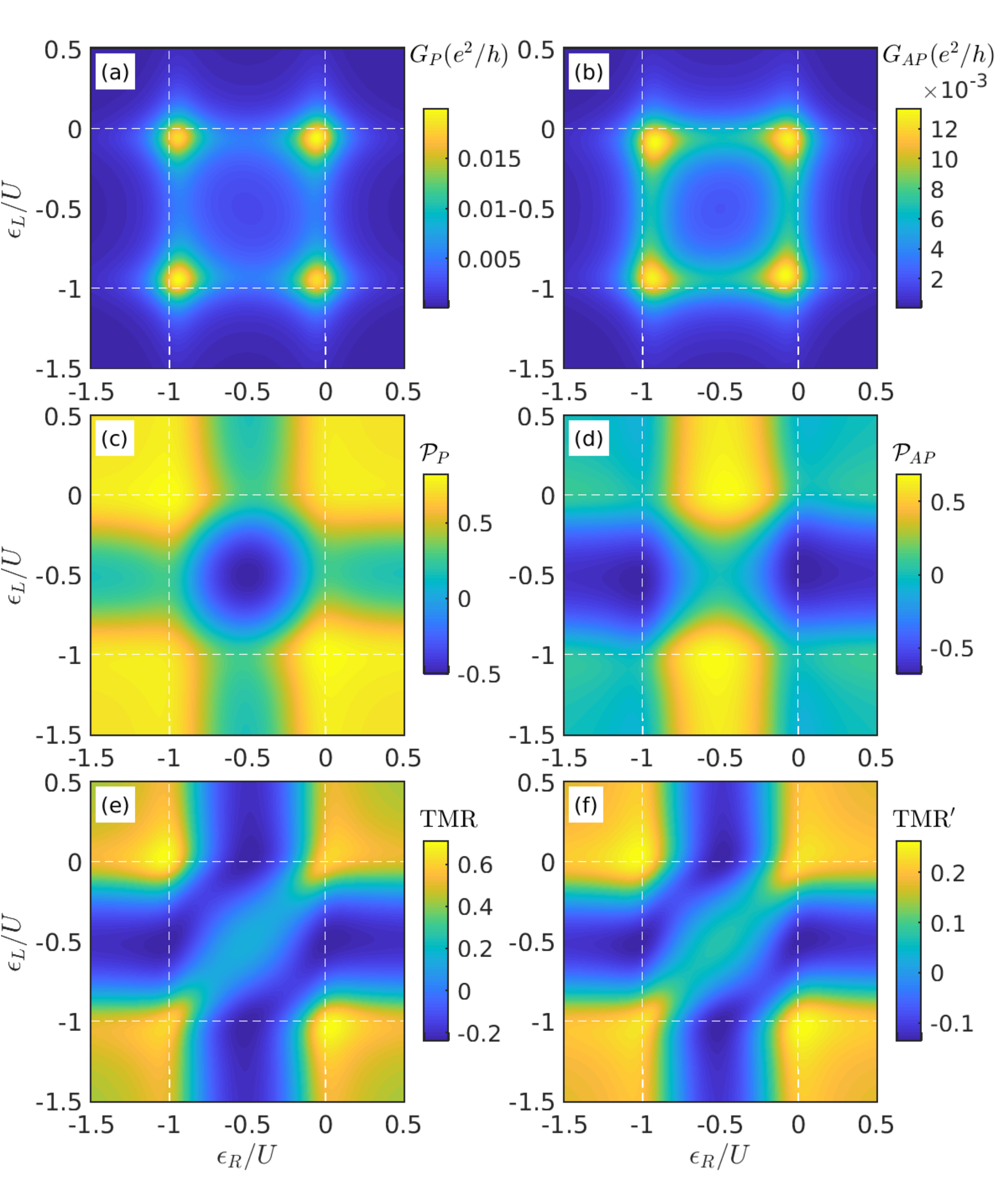}
	\caption{\label{Fig:FM2D10TK}
		The same as in \fig{Fig:FM2DTK} calculated for $T=10T_K$.
	}
\end{figure}

Let us now analyze the behavior of the spin polarization of the current
shown in Figs.~\ref{Fig:FM2DTK}(c) and (d) for the two magnetic configurations.
We note that the behavior for $\e_L=\e_R=-U/2$ has already been discussed
previously. The first general observation is that
the behavior of the spin polarization is completely different
depending on whether one of the dots is tuned to its
particle-hole symmetry point $\e_\alpha=-U/2$ or not.
In the case of parallel configuration,
when $\e_\alpha=-U/2$ holds for one of the dots,
the spin polarization is generally suppressed.
This can be understood by resorting to low-temperature estimations,
which, when assuming $\e_L=-U/2$, yield
\begin{equation} \label{Eq:PP}
	\Pol_P\approx \frac{(1-p) A_{R}^{\uparrow}(0) - (1+p)A_{R}^{\downarrow}(0) }
	{(1-p) A_{R}^{\uparrow}(0) + (1+p)A_{R}^{\downarrow}(0) }.
\end{equation}
It can be seen that the two terms in the numerator counterbalance each other
resulting in suppressed $\Pol_P$.
On the other hand, when $\e_\alpha\neq-U/2$ for the two dots,
the current is mainly determined by the majority spin channel,
which results in large spin polarization of the current,
reaching almost unity, see Fig.~\ref{Fig:FM2DTK}(c).
A completely different situation can be observed
for the antiparallel configuration, see Fig.~\ref{Fig:FM2DTK}(d).
Now, when $\e_\alpha\neq-U/2$, one finds suppressed spin polarization,
as the current is always proportional to a product of both
majority and minority spin spectral functions.
Nevertheless, when one of the dots exhibits the Kondo effect,
i.e. when $\e_\alpha=-U/2$ for one of the dots,
large spin polarization can be found in the system.
Moreover, the sign of $\Pol_{AP}$ turns out
to depend on whether  $\e_\alpha=-U/2$ for the left or
for the right quantum dot. Again, this behavior
can be understood assuming low-temperature limit.
Consider first the case of $\e_L=-U/2$, one then finds
\begin{equation} \label{Eq:PAP}
	\Pol_{AP} \approx \frac{(1-p) A_{R}^{\downarrow}(0) - (1+p)A_{R}^{\uparrow}(0) }
	{(1+p)A_{R}^{\uparrow}(0) + (1-p) A_{R}^{\downarrow}(0)  }.
\end{equation}
At first sight, this formula may seem similar to \eq{Eq:PP}.
However, there are important differences, as now the spin-up
(spin-down) component in the numerator is enhanced (suppressed)
by a $p$-dependent factor. Taking additionally into account the fact
that generally $A_{R}^{\uparrow}(0) > A_{R}^{\downarrow}(0)$,
one can see that $\Pol_{AP}$ becomes negative and reaches almost $-1$.
The formula for $\Pol_{AP}$ when $\e_R=-U/2$ can be obtained from
\eq{Eq:PAP} by replacing $R \leftrightarrow L$ and changing the sign.
Consequently, one then obtains $\Pol_{AP}\to 1$.
This behavior is clearly seen in Fig.~\ref{Fig:FM2DTK}(d).

Figure \ref{Fig:FM2D10TK} presents the
same quantities as in \fig{Fig:FM2DTK} calculated
for higher temperature, $T=10T_K$. Since at this temperature
the Kondo effect is smeared out by thermal fluctuations,
no cross-like feature is visible in the
conductance in the $(1,1)$ charge state regime,
irrespective of the system's magnetic configuration,
see Figs.~\ref{Fig:FM2D10TK}(a) and (b).
Instead, only the resonances at the corners of
this blockade regime are present.
As can be seen, basically all the features discussed in
the case of $T=T_K$ and visible in the behavior of
spin polarization and tunnel magnetoresistance
are also present for $T=10 T_K$. The main difference is  that
the most characteristic transport regimes are now broadened.
In particular, in the case of parallel configuration,
one finds enhanced $\Pol_P$ only when
each dot is evenly occupied, see \fig{Fig:FM2DTK}(c).
When one of the dots is singly occupied while
the second dot's occupation is even, $\Pol_P$ is reduced,
whereas in the $(1,1)$ blockade regime
the spin polarization becomes negative.
An exactly opposite behavior can be seen for $\Pol_{AP}$
[\fig{Fig:FM2DTK}(d)]. Now, the spin polarization
is suppressed whenever the total occupancy of the double dot is even,
whereas for odd total occupation, one finds
either $\Pol_{AP} \to 1$ or $\Pol_{AP} \to -1$,
depending  on which dot is singly occupied,
as already discussed previously.
On the other hand, in the case of TMR one can
observe that for higher temperatures
the two definitions give qualitatively similar results,
cf. Figs.~\ref{Fig:FM2D10TK}(e) and (f),
but of course there are still large quantitative differences.
Positive magnetoresistance develops only
when each of the dots has even occupancy, whereas
in the other transport regimes, the TMR becomes negative
except for a region along $\e_L=\e_R$ in the $(1,1)$ occupation regime.

\section{Non-linear response regime}

\begin{figure}[t]
	\includegraphics[width=1\columnwidth]{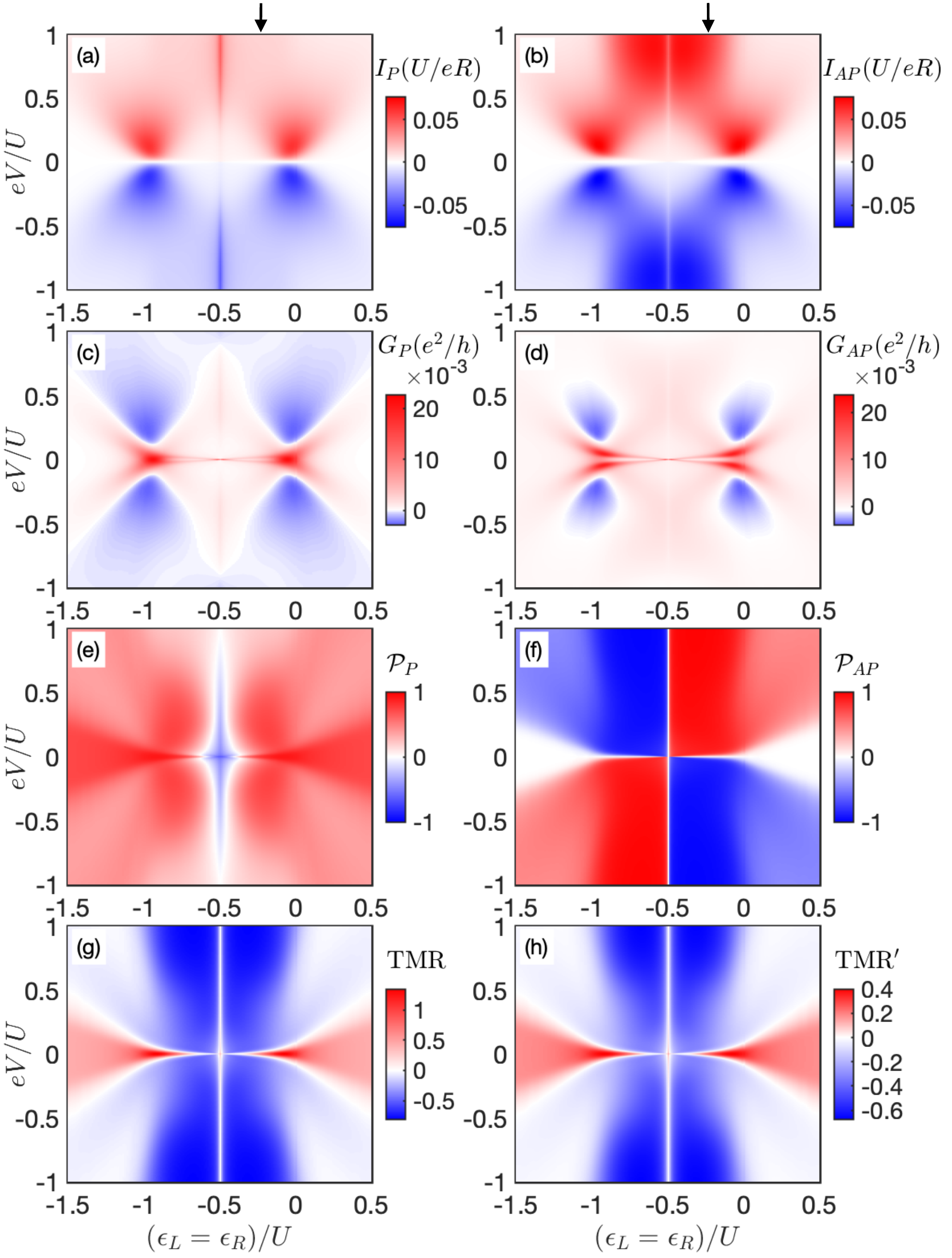}
	\caption{\label{Fig:2D}
		(a,b) The current, (c,d) differential conductance,
		and (e,f) current spin polarization
		in the (left column) parallel and (right column) antiparallel
		configuration as well as (g,h) the tunnel magnetoresistance
		as a function of bias voltage $V$ and position of the quantum dot levels $\e_L=\e_R$.
		The current is expressed in units of $U/eR$, where
		$R = R_\up + R_\down$ is the resistance of the junction
		between the two dots in the case of nonmagnetic leads.
		The arrows indicate the value of $\e_L=\e_R$ for which
		the spectral functions in Figs.~\ref{Fig:1}(b) and (c) are presented.
		The parameters are the same as in \fig{Fig:FM2DTK} with $T=0$.
	}
\end{figure}

We now turn to the discussion of nonequilibrium transport properties
and demonstrate that, in addition to the gate control discussed
in previous section, the operation of the considered double quantum dot
spin valve can be also tuned by applying the bias voltage.
The current flowing through the system at finite bias voltage
can be found from \eq{Eq:I}, whereas the tunnel magnetoresistance
and the current spin polarization can be evaluated from
Eqs.~(\ref{eq:TMR})-(\ref{eq:Pol}) by replacing the
linear conductance with the nonequilibrium current.
In the nonlinear response regime, for left-right symmetric
systems, the current satisfies the following relation
$I(V,\delta_L,\delta_R) = - I(-V,-\delta_L,-\delta_R)$,
where $\delta_\alpha = 2\e_\alpha + U$ denotes the
detuning from the particle-hole symmetry point of a given quantum dot.
On the other hand, for the other quantities
one has,
$X(V,\delta_L,\delta_R) = X(-V,-\delta_L,-\delta_R)$,
where $X=\{G,\Pol,{\rm TMR},{\rm TMR}^\prime \}$.

\begin{figure}[t]
	\includegraphics[width=1\columnwidth]{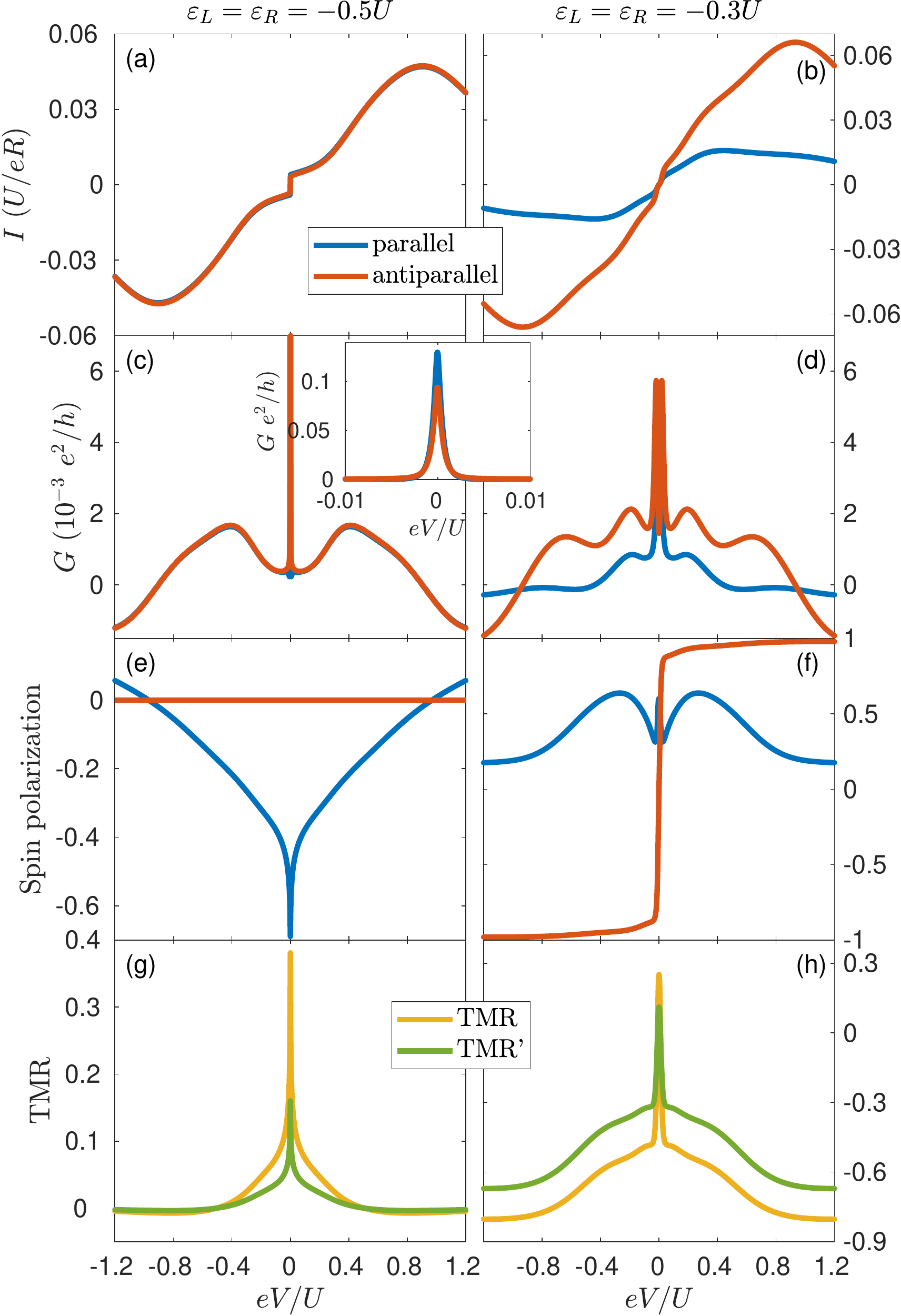}
	\caption{\label{Fig:1D}
		(a,b) The current, (c,d) differential conductance and
		(e,f) current spin polarization in the parallel (blue) and antiparallel (red)
		magnetic configuration as well as (g,h) the tunnel magnetoresistance
		as a function of the bias voltage $V$ calculated for (left column) $\e_L=\e_R=-0.5U$
		and (right column) $\e_L=\e_R=-0.3U$.
		The inset in (c) present a closeup of the
		differential conductance around the zero bias.
		The other parameters are the same as in \fig{Fig:2D}.
	}
\end{figure}

The current, differential conductance, current spin polarization
and tunnel magnetoresistance as a function of bias voltage and
the position of quantum dot levels $\e_L=\e_R$ are shown in \fig{Fig:2D}.
The current is plotted in units of $U/eR$, where $R = R_\up + R_\down$
is the resistance of the tunnel junction between the quantum dots
in the case of nonmagnetic leads.
The relevant cross-sections of the density maps are shown
in \fig{Fig:1D} for two selected values of the quantum dot level positions:
$\e_L=\e_R=-0.5U$ (left column) and $\e_L=\e_R=-0.3 U$ (right column),
where the latter value corresponds to the case
shown schematically in Figs.~\ref{Fig:1}(b) and (c).

When the dot levels are kept equal, i.e. $\e_L = \e_R$,
the exchange fields on each dot have the same magnitude and sign.
Then, the change of system transport properties can be obtained
by flipping the magnetization of one of the contacts
and changing the magnetic configuration from parallel to antiparallel one.
Such a situation is presented schematically in Figs.~\ref{Fig:1} (b) and (c),
where the spectral functions for $\e_L=\e_R = -0.3U$ are sketched.
For $\e_L=\e_R > -U/2$, $\exch <0$, such that each dot is
occupied by a spin-up electron at equilibrium and the spectral function
below the Fermi energy is mainly due to the spin-up component.
In this case, when a finite bias is applied to the system,
the current flowing through the device is very much suppressed at low temperatures.
This is due to the mismatch in the densities of states:
the occupied spin-up states from the left side do not have
available states on the right side, since there the empty states
are the spin-down ones, see Fig.~\ref{Fig:1} (b).
To open the system and enable a considerable current to flow,
one needs to change the magnetic configuration
of the device. When the magnetization of the right electrode
is flipped, the available states on the right are the spin-up states,
see Fig.~\ref{Fig:1} (c),
and a relatively large current can flow through the system.
The difference in the currents in both magnetic configurations is nicely visible in
Figs.~\ref{Fig:2D}(a) and (b). Generally,
$|I_P|<|I_{AP}|$ except for the transport regime where both dots
are evenly occupied (then the role of exchange field is diminished) and the voltage is sufficiently low.
The difference in currents results in an inverse tunnel
magnetoresistance, which develops in the regime of $-U<\e_\alpha <0$, whereas
for $\e_\alpha >0$ or $\e_\alpha <-U$, the magnetoresistance becomes positive,
see Figs.~\ref{Fig:2D}(g) and (h).

Because the density of states of either side of the system
is strongly energy- and spin-dependent, we observe that
the current displays a nonmonotonic behavior,
which is revealed in the negative differential conductance
visible in Figs.~\ref{Fig:2D}(c) and (d)
for the parallel and antiparallel magnetic configuration, respectively.
At low bias voltages, one can observe signatures of the Kondo resonance,
which is split due to the presence of exchange field,
see the red sharp feature at low bias in Fig.~\ref{Fig:2D}(d).
Such split Kondo resonance has already been observed
experimentally for single quantum dots \cite{Pasupathy2004Oct,Hauptmann2008Mar,Gaass2011Oct}.
Note that the differential conductance
exhibits a maximum at zero bias for $\e_L=\e_R=-U/2$
due to the Kondo effect, where its low-temperature value
is given by Eqs.~(\ref{Eq:GP}) and (\ref{Eq:GAP}).
To demonstrate the behavior of conductance out of this special point,
these maxima are not shown in \fig{Fig:2D}, since they are much
larger than the presented scale.
These Kondo resonances are however nicely visible in the
cross-sections shown in \fig{Fig:1D}(c) and the corresponding inset.

Let us now have a look at the specific cross-sections of \fig{Fig:2D}
and let us start with the case of $\e_L = \e_R = -0.5 U $ presented in the
left column of \fig{Fig:1D}. First we note that analytical formulas
for the equilibrium values of the transport quantities have been presented
in previous section. For $\e_L = \e_R = -0.5 U $, the exchange field is not effective
and the density of states is symmetric around the Fermi level
with a pronounced Kondo resonance. Because of that,
we observe moderate spin-resolved behavior: the currents
are comparable for larger voltages, resulting in vanishing tunnel magnetoresistance,
see \fig{Fig:1D}(g). The main difference between the magnetic configurations
is in fact visible at low bias, due to the different height of the Kondo
resonance in both spin channels, which is then  weakened as the voltage increases.
As mentioned in previous section, the low bias behavior of the TMR
for the DQD junction when $\e_L = \e_R = -0.5 U$ is similar to that of a single
magnetic tunnel junction.

The situation however drastically changes when the dots' levels are detuned from
the particle-hole symmetry point. The transport characteristics for this
situation are presented in the right column of \fig{Fig:1D}. Note that these
results correspond exactly to the situations sketched in Figs.~\ref{Fig:1}(b) and (c).
Now, due to the reasons discussed above, the current in the antiparallel
configuration is much larger than that in the parallel configuration, see \fig{Fig:1D}(b).
Moreover, as can be seen in \fig{Fig:1D}(d), the Kondo resonance is now suppressed and split.
The difference in currents is revealed in the magnetoresistance,
which quickly drops and changes sign as the voltage is increased.
In fact, for larger voltages, $|eV|\gtrsim U/2$, a
pronounced negative TMR develops in the system, revealing a considerable spin-valve behavior.
Interestingly, one now also observes a very high spin polarization of the flowing current
in the antiparallel configuration, reaching almost unity. The mechanism
responsible for this large spin polarization is associated with high
spin asymmetry in the density of states, as discussed above
and shown schematically in \fig{Fig:1}.

\begin{figure}[t!]
	\includegraphics[width=1\columnwidth]{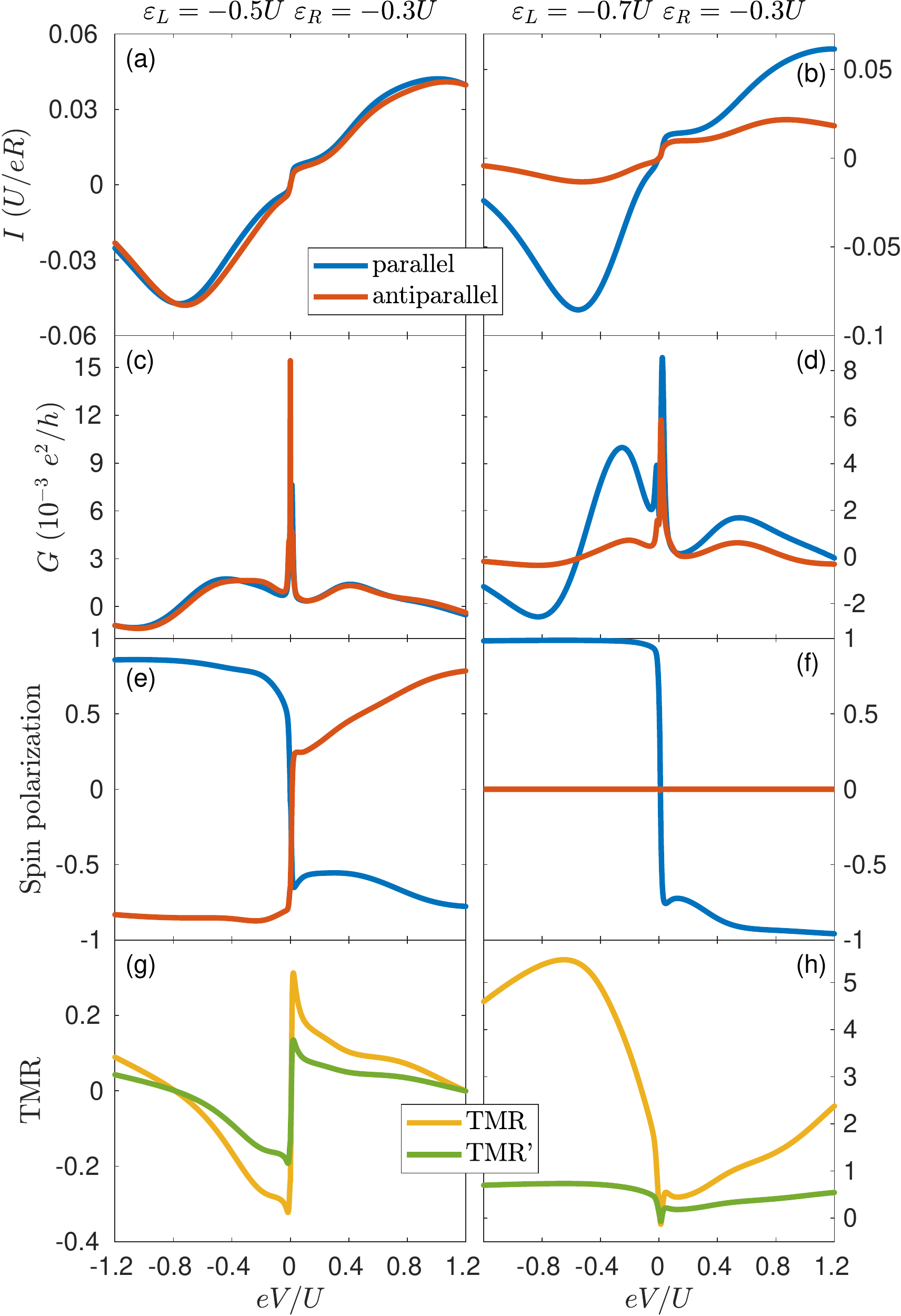}
	\caption{\label{Fig:1D2}
		(a,b) The current, (c,d) differential conductance and
		(e,f) current spin polarization in the parallel (blue) and antiparallel (red)
		magnetic configuration as well as (g,h) the tunnel magnetoresistance
		as a function of the bias voltage $V$ calculated for $\e_R=-0.3U$
		and (left column) $\e_L=-0.5U$ and (right column) $\e_L=-0.7U$.		
		The other parameters are the same as in \fig{Fig:2D}.
	}
\end{figure}

Another way of controlling the spin-resolved properties of the considered spin valve
is to tune the gate voltages and thus to adjust separately for each dot
the local magnetic fields of exchange-field origin.
This case is presented in \fig{Fig:1D2},
where we focus on the situation when the sign of the exchange field
on the right dot is fixed, i.e. $\exchR < 0$, favoring the spin-up (spin-down)
electrons in the parallel (antiparallel) configuration. Note that
the other situations can be obtained by invoking the symmetry of the current
with respect to the particle-hole symmetry point and the bias reversal,
$I(V,\delta_L,\delta_R) = - I(-V,-\delta_L,-\delta_R)$.
The left column of the figure presents the case when the exchange field
is present only in one of the dots, while the other dot is
tuned to the symmetry point. Even in this case
the exchange field in one dot gives rise to enhanced
spin polarization, which changes sign with either the bias reversal
or the flip of magnetic configuration, see \fig{Fig:1D2}(e).
However, the spin-valve behavior is rather moderate,
with low values of tunnel magnetoresistance, see \fig{Fig:1D2}(g).
To increase the difference in the current when the magnetic configuration
of the system is varied, one needs to induce
the spin splitting in both quantum dots---this case is presented
in the right column of \fig{Fig:1D2},
which corresponds to the situation when the  exchange fields have
different signs, but the same magnitude.
Due to the symmetry of densities of states,
the current in the antiparallel configuration is now not spin polarized.
However, a perfect current spin polarization is obtained
in the case of parallel configuration, which is visible especially for
negative bias voltage, see \fig{Fig:1D2}(f).
This is just contrary to the case presented
in \fig{Fig:1D}(f), where large spin polarization was found in the
antiparallel configuration. This demonstrates extremely high tunability
of the device, obtained either by changing the alignment of magnetic moments
of the leads or by tuning the level positions of the dots.
It is now also clearly evident that if the exchange fields
on both dots have different sign, an enhanced magnetoresistance
can be generated in the system, see \fig{Fig:1D2}(h).
Moreover, since the case presented in the left (right) column of 
\fig{Fig:1D2} corresponds to the situation
when the exchange field is present only in one of the dots (in both dots),
comparing the two cases reveals advantages of the double quantum dot junction 
over a junction containing a single quantum dot.

\begin{figure}[t!]
	\includegraphics[width=1\columnwidth]{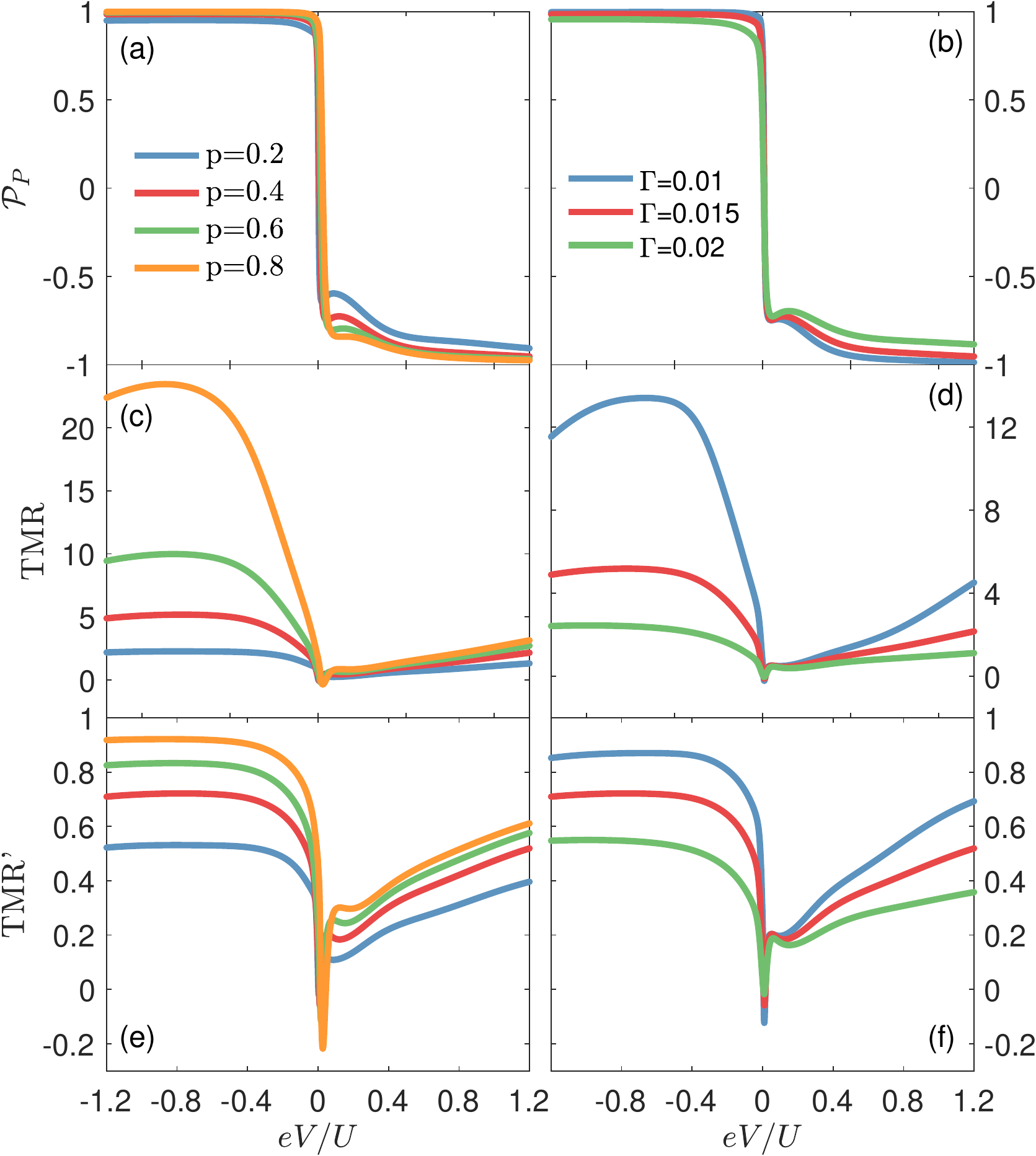}
	\caption{\label{Fig:1Dp}
		(a,b) The current spin polarization in the parallel
		magnetic configuration and the tunnel magnetoresistance
		(c,d) TMR and (e,f) TMR' as a function of the bias voltage $V$ calculated for $\e_L=-0.7U$
		and $\e_R=-0.3U$ for different values of (left column) spin polarization
		and $\Gamma = 0.015$ and (right column) for different values of
		coupling strength $\Gamma$ and $p=0.4$.
		The other parameters are the same as in \fig{Fig:2D}.
	}
\end{figure}

Finally, we analyze the effect of different values of spin polarization
as well as the impact of changing the coupling strength $\Gamma$
on the spin-valve behavior of the system.
The former case is presented in the left column of \fig{Fig:1Dp},
while the latter case is displayed in the right column of this figure.
We now only focus on the spin-resolved transport properties,
i.e. on the current spin polarization and the tunnel magnetoresistance.
Moreover, since $\Pol_{AP}$ vanishes for the chosen
set of level positions, we only present the behavior of $\Pol_P$.
One can see that the current spin polarization
is very high also for smaller values of $p$
(especially for negative bias),
and it approaches exactly unity with increasing $p$, see \fig{Fig:1Dp}(a).
The change of degree of the contacts' spin polarization
has larger effect on the tunnel magnetoresistance.
One can see that the higher $p$, the larger
TMR the system exhibits. This is related with
the fact that increasing $p$ results in an enhancement
of exchange field and, consequently, the magnetoresistive
properties are also enlarged, see Figs.~\ref{Fig:1Dp}(c) and (e).
On the other  hand, if the coupling
strength is changed while the spin polarization $p$ is constant,
one observes a strong impact on the behavior of the TMR,
while $\Pol_P$ rather weakly depends on the considered
values of $\Gamma$ (note that for negative bias the spin polarization is already
essentially full anyway), see \fig{Fig:1Dp}(a) and (b).
In particular, the tunnel magnetoresistance
for negative bias increases when $\Gamma$ is lowered.
This can be understood by realizing that for smaller couplings
the width of resonant peaks in the local density of states decreases
and their influence becomes suppressed. As a result,
the difference between the opposite spin components
of the spectral function in the singly occupied regime becomes enhanced,
yielding increased magnetoresistance, see Figs.~\ref{Fig:1Dp}(d) and (f).


\section{Conclusions} \label{conclusions}

We have analyzed the transport properties of a double quantum
dot based spin valve, with each dot strongly coupled to its own ferromagnetic lead
and weakly attached to each other.
The current flowing through the system has been calculated perturbatively in
the hopping between the dots, while the properties
of a quantum dot-ferromagnetic lead subsystem have been
determined by using the numerical renormalization group method.
This allowed us to capture all the electron correlations in an essentially
exact manner. At low temperatures, the local density of states
of the quantum dot-lead subsystem exhibits the Kondo resonance,
which however becomes suppressed by the exchange field
that splits the dot level when it is detuned from the particle-hole symmetry point.
Consequently, by tuning the position of the quantum dot levels,
one can adjust both the magnitude and sign of the spin splitting
of the energy levels of the double dot.
This enables a large voltage tunability of the device.

In the linear response regime, we have demonstrated
that the tunnel magnetoresistance, associated with
the change of magnetic configuration from the parallel
to the antiparallel one, can reach large positive or negative values
depending on the positions of the quantum dot levels.
Moreover, we have also identified transport regimes
of enhanced spin polarization of the linear conductance,
which can be tuned by either changing the position
of double dot levels or by switching the magnetic configuration of the device.

Finite bias voltage applied to the system
provides another means of controlling the behavior of
the considered spin valve. We have shown that
one can obtain a perfect spin polarization of nonequilibrium
current, whose sign can be controlled either by  the gate voltages
of by external magnetic field used to flip the magnetic configuration
of the system. We have also predicted large magnetoresistance
of the system in appropriate transport regime.

This work demonstrates that double quantum dots strongly attached to
external ferromagnetic contacts behave as voltage-tunable
spin valves with very prospective spin-resolved properties. The usage of
an  exchange field, which mimics magnetic field acting locally on a given quantum dot,
is crucial for obtaining enhanced magnetoresistance and current spin polarization.
It is beneficial over the usage of external magnetic field
explored recently \cite{Bordoloi2020Aug},
and allows to obtain much larger magnetoresisitive response of the system.


\begin{acknowledgments}
This work was supported by the Polish National Science
Centre from funds awarded through the decision No. 2017/27/B/ST3/00621.
S.C. acknowledges funding from SuperTop QuantERA network,
the Ministry of Innovation and Technology
and the NKFIH within the Quantum Information National Laboratory of Hungary,
and the Quantum Technology National Excellence Program
(Project Nr. 2017-1.2.1-NKP-122017-00001). 
\end{acknowledgments}



%

\end{document}